\documentclass[aps,prl,reprint,superscriptaddress,floatfix,amsmath,boldmath]{revtex4-1} 

\usepackage{xcolor}
\usepackage[normalem]{ulem}
\usepackage{graphicx}

\definecolor{g-blue}{rgb}{0.83,0.95,1}
\definecolor{g-yellow}{rgb}{1,1,0.7}
\definecolor{g-green}{rgb}{0.9,1,0.9}
\definecolor{green}{rgb}{0,0.6,0}
\definecolor{cyan}{rgb}{0,0.7,0.7}
\definecolor{black}{rgb}{0,0,0}
\definecolor{grey}{rgb}{0.4 ,0.4 ,0.4 }

\usepackage{bm}
\usepackage{amssymb}
\usepackage{amsfonts}
\usepackage{amsmath}
\usepackage{graphicx}
\usepackage{amsmath,bm,epsfig}

\def \ed {\end{document}}
\def\Fbox#1{\vskip1ex\hbox to 8.5cm{\hfil\fboxsep0.3cm\fbox{%
\parbox{8.0cm}{#1}}\hfil}\vskip1ex\noindent}  

\newcommand{\Eq}[1]{Eq.\,(\ref{#1})}
\newcommand{\Eqs}[1]{Eqs.\,(\ref{#1})}
\newcommand{\Ref}[1]{Ref.\,\cite{#1}}

\newcommand{\C}[1]{{\mathcal{#1}}}    

\renewcommand{\sb}[1]{_{\text {#1}}}  

\begin{document}
	
	\title{From Kinetic Instability to Bose-Einstein Condensation and   Magnon Supercurrents}
		
	\author{Alexander~J.~E.~Kreil}	
	\email{kreil@rhrk.uni-kl.de}
	\affiliation{Fachbereich Physik and Landesforschungszentrum OPTIMAS, Technische Universit\"at Kaiserslautern, 67663 Kaiserslautern, Germany \looseness=-1}
		
	\author{Dmytro~A.~Bozhko}
	\affiliation{Fachbereich Physik and Landesforschungszentrum OPTIMAS, Technische Universit\"at Kaiserslautern, 67663 Kaiserslautern, Germany \looseness=-1}
	
	\author{Halyna~Yu.~Musiienko-Shmarova}
	\affiliation{Fachbereich Physik and Landesforschungszentrum OPTIMAS, Technische Universit\"at Kaiserslautern, 67663 Kaiserslautern, Germany \looseness=-1}
	
	\author{Victor~S.~L'vov}
	\affiliation{Department of Chemical and Biological Physics, Weizmann Institute of Science, Rehovot 76100, Israel \looseness=-1}
	
	\author{Anna~Pomyalov}
	\affiliation{Department of Chemical and Biological Physics, Weizmann Institute of Science, Rehovot 76100, Israel \looseness=-1}
	
	\author{Burkard~Hillebrands}
	\affiliation{Fachbereich Physik and Landesforschungszentrum OPTIMAS, Technische Universit\"at Kaiserslautern, 67663 Kaiserslautern, Germany \looseness=-1}
	
	\author{Alexander~A.~Serga}
	\affiliation{Fachbereich Physik and Landesforschungszentrum OPTIMAS, Technische Universit\"at Kaiserslautern, 67663 Kaiserslautern, Germany \looseness=-1}
	

\begin{abstract}
Evolution of an overpopulated gas of magnons to a Bose-Einstein condensate and excitation of a magnon supercurrent, propelled by a phase gradient in the condensate wave function, can be observed at room-temperature by means of the Brillouin light scattering spectroscopy in an yttrium iron garnet material. We study these phenomena in a wide range of external magnetic fields in order to understand their properties when externally pumped magnons are transferred towards the condensed state via two distinct channels: A multistage Kolmogorov-Zakharov cascade of the weak-wave turbulence or a one-step kinetic-instability process. Our main result is that opening the kinetic instability channel leads to the formation of a much denser magnon condensate and to a stronger magnon supercurrent compared to the cascade mechanism alone. 
\end{abstract}

\maketitle
	
Bose-Einstein condensation (BEC) is a fascinating quantum phenomenon that manifests itself in the formation of a coherent macroscopic state from the chaotic motions in a thermalized many-particle system. In spite of being a consequence of equilibrium Bose statistics \cite{Einstein1924,Einstein1925}, BEC can also occur in rather non-equilibrium systems, such as overpopulated gases of bosonic quasiparticles -- excitons \cite{Butov2001}, polaritons \cite{Kasprzak2006,Balili2007,Rodriguez2013}, photons \cite{Klaers2010}, and magnons \cite{Bunkov1984,Demokritov2006} -- as a result of local quasi-equilibrium conditions near the bottoms of their frequency spectra. BEC formation in different quasiparticle systems constitutes a challenge of fundamental importance for physics in general and for possible applications in which a BEC of quasiparticles is used for data processing. 
A particularly interesting case is given by a magnonic BEC \cite{Demokritov2006, Serga2014} observed at room temperature in the low-damping ferrimagnetic material yttrium iron garnet (YIG, $\mathrm{Y_{3}Fe_{5}O_{12}}$) \cite{Cherepanov1993}. Such a condensate is created from a magnon gas overpopulated by intensive parametric injection of magnons. The parametric magnons are then transferred by step-by-step cascade processes \cite{ZLF,Demidov2008,Kopietz2012,Bozhko2015} down the frequency band, followed by a thermalization of low-energy magnons into the BEC  state \cite{Serga2014,Clausen2015}. Under certain conditions, the cascade processes can be augmented by a direct transfer of the parametrically injected magnons to the lowest energy states. In this case, referred to as a kinetic instability (KI) process \cite{Lavrinenko1981,Melkov1991}, a dense cloud of non-coherent magnons is formed close to the BEC point. By the energy conservation law, the same number of parametric magnons is transferred to  higher energy states and, thus, a strongly non-equilibrium magnon gas distribution,  characterized by two population maxima, is formed. \looseness=-1

In this Letter, we show experimentally and theoretically that KI provides favorable conditions for a more efficient magnon condensation and for a stronger BEC-related supercurrent spin transport \cite{Bozhko2016} compared to the cascade-only scenario. \looseness=-1

In our experiment, BEC formation is initiated by an external quasi-homogeneous electromagnetic field of frequency $\omega_\mathrm{p}$ that parametrically excites pairs of magnons with wavevectors $\pm \mathbf{q}_\mathrm{p}$  and frequencies $\omega(\pm \mathbf{q}_\mathrm{p})$ via the three-wave decay process \cite{NSW,Gurevich1996,Vasyuchka2013}: $\omega_\mathrm{p} \Rightarrow \omega(\mathbf{q}_\mathrm{p})+ \omega(-\mathbf{q}_\mathrm{p})$.
At some threshold pumping power, the parametric magnon excitation  compensates the natural magnon damping. A  magnon mode, that has the lowest damping frequency $\gamma(\mathbf{q}_\mathrm{p})$ and the strongest coupling to the pumping field, starts growing exponentially in time. \looseness=-1

\begin{figure}[b]
	\includegraphics[width=0.98\columnwidth]{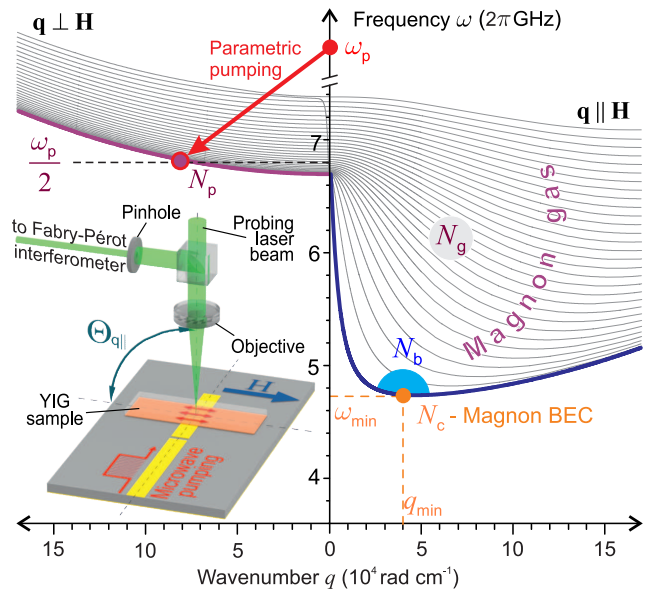}
	\caption{
		\label{F:1} Magnon spectrum of a 5.6\,$\mu$m thick YIG film magnetized in plane by a bias magnetic field $H=1700$\,Oe shown for the wavevector $\textbf{q}$ perpendicular (left part) and parallel (right part) to the applied field. The red arrow illustrates the process of parametric pumping. $N_\mathrm{p}$ - total number of parametrically excited magnons at $\omega_\mathrm{p}/2$; $N_\mathrm{c}$ - number of BEC magnons at $\omega_\mathrm{c} = \omega_\text{min}$; $N_\mathrm{b}$ -- number of gaseous  magnons near $\omega_\mathrm{min}$ and $q_\mathrm{min}$;  $N_\mathrm{g}$ -- number of magnons in the parametrically overpopulated gas of magnons below $\omega_\mathrm{p}/2$. The inset shows a sketch of the experimental setup. By using a resonance microstrip circuit, a pumping electromagnetic wave is delivered to the YIG sample and conditions of parallel parametric pumping are realized. The light inelastically scattered by magnons is analyzed by a Fabry-P\'{e}rot interferometer.  Wavenumber-selective probing of magnons with wavevectors $\textbf{q} \parallel \textbf{H}$ is realized by varying the incidence angle $\Theta_{q_\parallel}$ between the field $\textbf{H}$ and the probing laser beam \cite{Sandweg2010}.
	}
\end{figure}

Figure\,\ref{F:1} illustrates the frequency spectrum $\omega(\mathbf{q})$ of magnons in a YIG film magnetized in plane by a bias magnetic field $\mathbf{H}$. 
The  minimum of the spectrum is located at some $\mathbf{q}=\pm \mathbf{q}_{\mathrm{min}}$, where $\mathbf{q}_{\mathrm{min}} \parallel \mathbf{H}$. The parametrically pumped  magnons are excited with $\mathbf{q}_\mathrm{p} \perp \mathbf{H}$  and fill a part of the iso-frequency surface $\omega(\mathbf{q}_\mathrm{p})=\omega_\mathrm{p}/2$ \cite{NSW, Serga2014,Clausen2015} above the bottom of the frequency spectrum $\omega_{\mathrm{min}}=\min_\mathbf{q}\{\omega(\mathbf{q})\}=\omega(\mathbf{q}_{\mathrm{min}})$. It can be shown that nonlinear magnon scattering causes their transfer from the injection area with the frequency $\omega_\mathrm{p}/2$ toward the bottom frequency area around $\omega_{\mathrm{min}}$ \cite{Bun}. Afterwards, thermalization of the magnon occupation number distribution $n(\mathbf{q})$, local in phase space, can lead to the BEC of magnons at $\mathbf{q}=\mathbf{q}_{\mathrm{min}}$ \cite{Demokritov2006}. \looseness=-1

There are two possible channels of the magnon transfer toward $\omega_{\mathrm{min}}$. The first one is a step-by-step flux of $n(\mathbf{q})$ \cite{ZLF} (Kolmogorov-Zakharov cascade) similar to the Richardson-Kolmogorov cascade of the turbulent kinetic energy in classical hydrodynamic turbulence \cite{Fri}. The second channel can arise from a specific four-magnon scattering process \cite{Lavrinenko1981,Melkov1991} that is the fusion of two parametrically injected magnons with $\omega(\mathbf{q}_1)\approx\omega(\mathbf{q}_2)\approx \omega_\mathrm{p}/2$ resulting in the creation of a secondary bottom magnon with $\omega(\mathbf{q}_3) \gtrsim \omega_{\mathrm{min}}$ and a secondary high-frequency magnon with $\omega(\mathbf{q}_4) > \omega_\mathrm{p}/2$ in the $2\Leftrightarrow2$ scattering process determined by the conservation laws: \looseness=-1
\begin{subequations}\label{KIC}
 \begin{eqnarray}\label{KICa}
  \omega(\mathbf{q}_1)+ \omega(\mathbf{q}_2) &=&  \omega(\mathbf{q}_3)+\omega(\mathbf{q}_4)\,, \\ \label{KICb}
  \mathbf{q}_1 + \mathbf{q}_2 &=& \mathbf{q}_3+ \mathbf{q}_4\,, \\ \label{KI Cc}
  \omega(\mathbf{q}_1) \approx \omega(\mathbf{q}_2) \approx \omega_\mathrm{p}/2\,, && \omega(\mathbf{q}_3) \gtrsim \omega_{\mathrm{min}}\ .
\end{eqnarray}
\end{subequations}
In contrast to the dynamical process of the parametric instability, in which the strong phase correlation between three waves (photon and two magnons) plays a crucial role, the phase correlation  between  waves, involved in \Eq{KICa} is very weak \cite{NSW}. As a result, the magnon evolution in the KI process can be described in terms of their occupation numbers $n(\mathbf{q})$ in the framework of a kinetic equation \cite{ZLF,NSW}.
The processes \Eq{KICa} are allowed for some portion $\C{N}_\mathrm{p}$ of the total number of the parametrically pumped magnons $N_\mathrm{p}$. When  $\C{N}_\mathrm{p}$ exceeds the critical value $\C{N}_\mathrm{cr} = \sqrt{\gamma_\mathrm{b}/A}$, the modified damping frequency $\Gamma_\mathrm{b}= \gamma_\mathrm{b} - A \C{N}_\mathrm{p}^{\,2} $, where $\gamma_\mathrm{b}$ is the original damping frequency and $A$ is some dimensional parameter, becomes negative \cite{Lavrinenko1981} and the number of the gaseous bottom magnons $N_\mathrm{b}\propto \exp(-\Gamma_\mathrm{b} t)$ grows exponentially. This phenomenon, discovered in \Ref{Lavrinenko1981} and labeled there \emph{kinetic instability} (KI), is inherent to systems of nonlinear waves to the same extent as the parametric instability and the Bose-Einstein condensation processes.

\begin{figure*}
	 \includegraphics[width=17 cm]{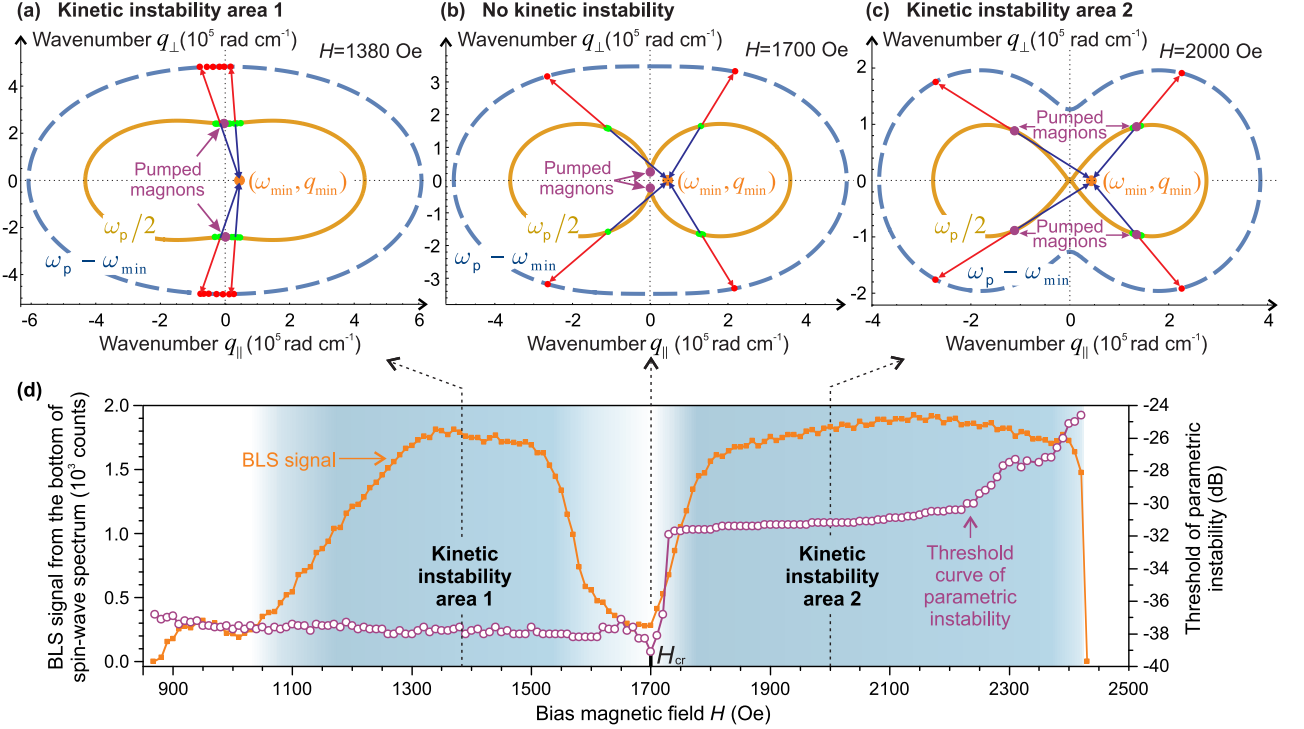}
	\caption{\label{F:2}
    Panels (a), (b) and (c): Diagrams describing the KI processes at different bias magnetic fields. The isofrequency curves $\omega(H,q_\parallel,q_\perp)=\omega_{\mathrm{p}}/2$ of pumped magnons are shown in each panel by brown solid lines, while isofrequency curves of the frequency $\omega(H,q_\parallel,q_\perp)=\omega_{\mathrm{p}} - \omega_\mathrm{min}$ are shown by blue dashed lines. The BEC's spectral position at $(\omega_\mathrm{min}, \mathbf{q}_\mathrm{min})$ is indicated by the orange dots. The initial magnon positions are marked  by small light green dots. The big magenta dots on the isofrequency curve $\omega_\mathrm{p}/2$ correspond to the position of parametrically pumped magnons with $\mathbf{q}_1\approx \mathbf{q}_2$, which can potentially participate in the KI processes.  The blue and red arrows illustrate the potentially possible magnon scattering processes with the spectral positions in the frequency band $\omega_\mathrm{p}/2 \pm 2 \pi \cdot 50$\,MHz. Panel (d): Measured BLS intensities of scattered magnons (orange squares) in comparison with the threshold curve of the parametric instability process (magenta empty circles). The shaded regions, in which KI processes are allowed by conservation laws, are marked as ``Kinetic instability areas 1 and 2'' \looseness=-1.}
\end{figure*}

In the following we compare the dynamics of the bottom magnons for different values of the external static bias magnetic field $H$, which determines the magnon frequency $\omega(H,\mathbf{q})$ in such a way that the KI channel is allowed or forbidden by the conservation laws Eq.\,\eqref{KIC}. In the latter case, only the Kolmogorov-Zakharov cascade is responsible for the magnon transfer to the bottom spectral area.
We show experimentally and discuss ana\-ly\-ti\-cal\-ly that the KI channel provides a much larger number of bottom magnons participating in the BEC formation than the cascade channel (in our conditions $N\sb b$  increases by more than an order of magnitude compared to the cascade-only case). 

In our experiments, the magnons are injected in a 5.6\,$\mu$m-thick in-plane magnetized YIG film at a frequency of $\omega_\mathrm{p}/2 =$ $2 \pi \cdot 6.8$\,GHz by pulsed parametric pumping with a peak power of 40\,W. The time evolution of the mag\-non density is studied at room temperature by frequency-, time- and wavevector-resolved BLS spectro\-scopy \cite{Serga2014,Buettner2000,Sandweg2010,Serga2012} (see a schematic view of the experimental setup in  Fig.\,\ref{F:1}). The detected BLS signal is proportional to the total number $N_\mathrm{tot} = N_\mathrm{b} + N_\mathrm{c}$ of gaseous and condensed magnons in the vicinity of ($\omega_\mathrm{min}, q_\mathrm{min}$).  

To understand the processes of magnon transfer to the bottom of the spectrum, in Figs.\,\ref{F:2}\,(a)--(c) we plot  the magnon scattering diagrams, representing the kinetic instability process for three different values of the bias magnetic field $H=1380\,$Oe (a), $1700\,$Oe (b) and $2000\,$Oe (c), in comparison with the intensity of the BLS signals from the bottom spectral area and with the threshold power of the parametric instability process, which are shown as functions of the bias magnetic field $H$ in Fig.\,\ref{F:2}\,(d). 

The KI processes is always allowed if the parametric magnons occupy entirely the isofrequency curves $\omega(H,q_\parallel,q_\perp)=\omega_\mathrm{p}/2$.  However, at different bias magne\-tic fields the parametric pumping process populates distinctly different parts of this curve (cf. Figs.\,\ref{F:2}\,(a)--(c)). The mag\-non frequency spectrum, shown in Fig.\,\ref{F:1}, depends on the bias magnetic field. With increasing $H$, the spectrum is shifted upward. At some critical field $H_\mathrm{cr} \approx 1700$\,Oe, the frequency of parametric magnons $\omega\sb p/2$ meets the minimum of the spectrum for $\mathbf{q} \perp H$. For these magnons with $\mathbf{q} \rightarrow 0$, the damping frequency $\gamma(\mathbf{q}_\mathrm{p})$ is lowest, making them most easily exited. This is clearly seen in Fig.\,\ref{F:2}(d): The parametric threshold curve has a pronounced minimum at the critical magnetic field $H_\mathrm{cr} \approx 1700$\,Oe. However, at these conditions, the KI process is strongly suppressed, as it is evident from the deep minimum of the BLS intensity curve in the same figure.

Let us consider first magnetic fields $H \leq H_\mathrm{cr}$, for which the iso\-frequency curve $\omega(H,q_\parallel,q_\perp)=\omega_\mathrm{p}/2$ (Fig.\,\ref{F:2}(a)) has a quasi-elliptical shape. Here, the most efficiently injected magnons have wavevectors $\mathbf{q}_\mathrm{p}=\pm \mathbf{q}_\perp \perp \mathbf{H}$ \cite{Serga2012,Zautkin1971}, overlapping with the spectral area of kinetic instability. As a result, the parametric magnons efficiently participate in the KI process and $\C{N}_\mathrm{p} \approx N_\mathrm{p}$.
At high pumping powers, typical for our experiments, the spectral area of parametrical excitation spreads towards polar angles $\Theta_\mathbf{q} = \measuredangle (\mathbf{q}, \mathbf{H}) \gtrsim 45^\circ$ \cite{Serga2014,NSW}. Therefore, the number of parametric magnons $\C{N}_\mathrm{p}$, for which the KI process is allowed by \Eqs{KIC}, has a wide maximum between 1000\,Oe and 1700\,Oe. As a result, the total number $N_\mathrm{tot}$ of the bottom and the BEC magnons, observed by BLS, has a broad hump in the corresponding magnetic field range, shaded as ``Kinetic instability area 1'', in Fig.\,\ref{F:2}(d). \looseness=-1

Next, we consider the bias magnetic fields around $H_\mathrm{cr}$ from 1650\,Oe to 1750\,Oe.
Here, the spectral positions of the parametric magnons are well separated in the phase space from the KI area (Fig.\,\ref{F:2}\,(b)). Therefore, the number $\C{N}_\mathrm{p}$ of parametric magnons with $\Theta_{\mathbf{q}} \gtrsim 45^\circ$, which can participate in the KI process, sharply decreases. The increment of the KI, $ \propto \C{N}_\mathrm{p}(H)^2 - \C{N}_\mathrm{cr}^2$, strongly decreases and can even become negative, meaning that at this range of $H$, the KI processes are either strongly suppressed or forbidden. \looseness=-1

Finally, for $H \geq H_\mathrm{cr}$ the form of the isofrequency line $\omega(H,q_\parallel,q_\perp)=\omega_\mathrm{p}/2$ changes topologically from almost elliptical shape to a ``$\infty$'' shape (cf. panels (a) and (c) in Fig.\,\ref{F:2}). In this case, the frequency $\omega_\mathrm{p}/2$ lies below the magnon branch with $\mathbf{q} \perp \mathbf{H}$ (cf. Fig.\,\ref{F:1}) and conditions of the parametric pumping essentially change \cite{Neumann2009}. As a result,  the parametric instability threshold sharply increases by about 6\,dB  (Fig.\,\ref{F:2}(d), magenta circles). It can be shown\,\cite{NSW} that for $H \geq H_\mathrm{cr}$ and for large pumping powers, the parametric magnons are excited along the entire isofrequency curve $\omega(\mathbf{q})=\omega_\mathrm{p}/2$. As a result, $ \C{N}_\mathrm{p} (H)$ has a broad maximum. This new area of KI is marked in Fig.\,\ref{F:2}(d) as ``Kinetic instability area 2''.

\begin{figure*}
 \includegraphics[width=17 cm]{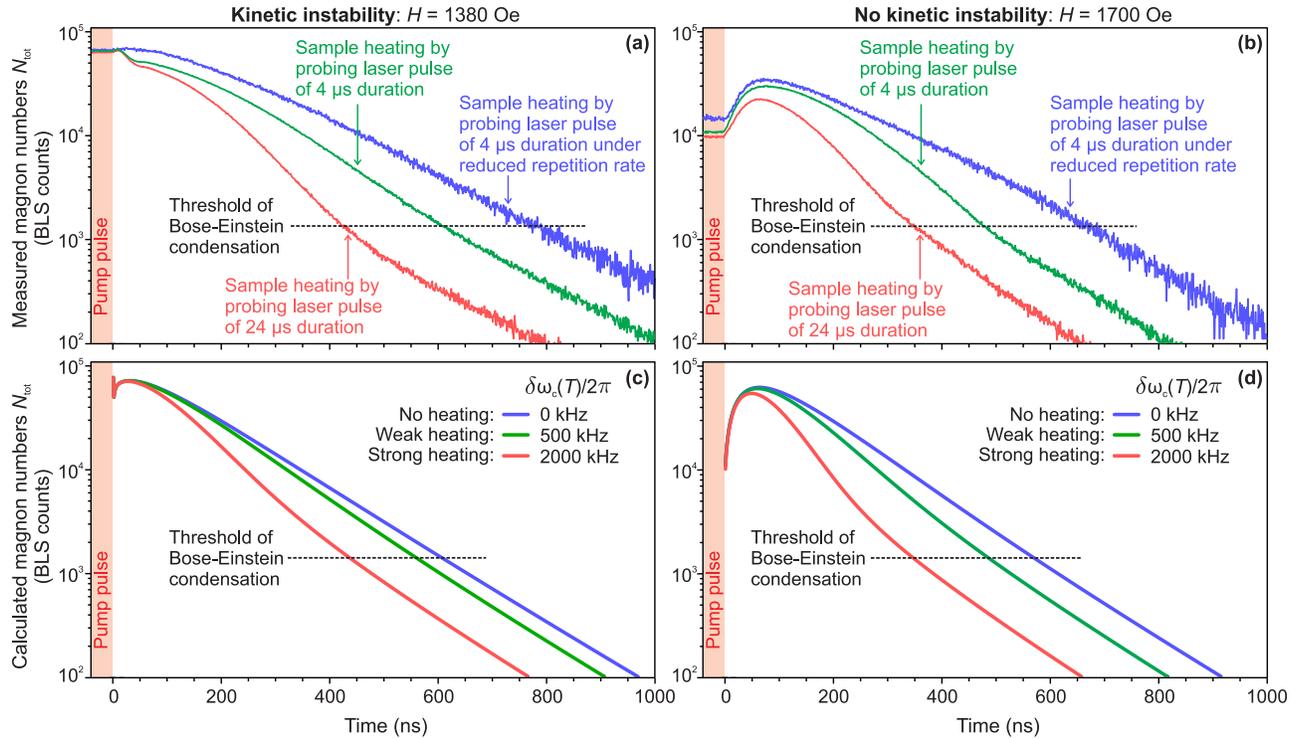}
	\caption{\label{F:3}
Experimental and calculated dynamics of the total number of the bottom magnons $N_\mathrm{tot}$ influenced by a local heating of the YIG film.
The BLS data are collected in a 150\,MHz frequency band near the bottom of the magnon spectrum for different heating regimes in the first area of kinetic instability (panel (a)) and at the critical bias field $H=H_\mathrm{c}$ (panel (b)). The red, green, and blue curves represent strong, medium, and negligibly small local heating, respectively. Panels (c) and (d): Theoretical dependencies of the magnon densities, calculated  taking into account different temperature-dependent shifts $\delta \omega_{\mathrm{c}} (T)$ of the BEC frequency using the model from \cite{Bozhko2016} with the initial conditions corresponding to the cases with and without KI. \looseness=-1
}
\end{figure*}

Note that in both KI regions, in the measured data the number of BLS counts $\propto N_\mathrm{tot}(H)$ is above 1000 and reaches 2000, while in the area around $H_\mathrm{cr}$ this number does not exceed 300 counts. Thus, we conclude that the KI channel is at least by an order of magnitude more efficient in the transfer of parametrically exited magnons to the BEC area than the step-by-step Kolmogorov-Zakharov cascade. 

The coherent condensed state, formed from the low-frequency magnons, may be evidenced by the detection of a magnon supercurrent excited by a local heating of the YIG film \cite{Bozhko2016}. Given the efficiency of the KI channel in creating the overpopulated magnon gas at the bottom frequency area, one can expect it to affect the properties of both the BEC and the supercurrent.
To study this influence, we compare the free evolution of a magnon BEC after the termination of the parametric pumping pulse under different local heating conditions.
To control the heating, we use a pulsed laser focused onto the sample. The sample is probed by the same laser light, which is synchronized with the microwave pumping pulses \cite{Bozhko2016}.
The temperature of the probing BLS point is changed by adjusting the duration and the repetition rate of probing laser pulses. \looseness=-1

The magnon decay dynamics is shown in Fig.\,\ref{F:3} for $H=1380\,$Oe, corresponding to the first area of KI, and for $H=1700\,$Oe, where KI is negligible.
The main differences in the initial stages of the BLS signal evolution between these cases is the total number of bottom magnons observed just before the termination of the pumping pulse: $N_{\mathrm{tot}} \simeq 10^5$ BLS counts with KI in Fig.\,\ref{F:3}(a) and $N_{\mathrm{tot}} \simeq 10^4$ BLS counts without KI in Fig.\,\ref{F:3}(b). However, in the latter case the BLS signal additionally increases in course of the pumping-free evolution of the magnon system.

To understand this difference, we note that the cascade process creates an almost uniform distribution of the thermalized magnons in a relatively wide region around the bottom of the magnon spectrum, as well as every where else in the magnon gas. After the pumping pulse ends, these magnons move to the minimum energy state by means of non-linear four-magnon scattering and form there a pronounced delayed peak of the magnon gas density \cite{Serga2014}, clearly visible in Fig.\,\ref{F:3}(b). In contrast, the KI process strongly populates spectral states near $\omega \sb{min}$ already in the course of action of the pumping pulse, as it is observed by BLS. The following redistribution of these magnons between the gaseous and BEC states does not change $N_\mathrm{tot}$ and, thus, is not reflected in the BLS dynamics. In Fig.\,\ref{F:3}(a), a barely visible after-pulse signal increase may come from a minor contribution from far-tails of the initial density peak around $\omega_\mathrm{min}$.

Nevertheless, with or without KI, the heating of the probing point by a probing laser pulse leads to an enhanced signal decay at high magnon densities $N_\mathrm{tot}(t)$. This phenomenon, understood as a supercurrent-related outflow of condensed magnons from the heated probing spot \cite{Bozhko2016}, serves as a signature of the spontaneously established coherent magnon phase -- the magnon BEC, independent of the magnon transfer scenario.
Figures\,\ref{F:3}(a) and \ref{F:3}(b) show that the supercurrent is not induced in experiments with a space-homogeneous cold film (reduced repetition rate, short laser pulse of 4\,$\mu$s duration, blue lines), become noticeable for moderate local heating (high repetition rate, short laser pulse, green lines) and is pronounced for strong heating (high repetition rate, long laser pulse  of 24\,$\mu$s duration, red lines). At the later stage of the decay, when the BEC already disappeared and no phase coherency can be assumed in the magnon gas, the supercurrent vanishes in all cases.
Therefore, with and without KI, the  BLS signal decay in the inhomogeneously heated YIG film clearly indicates the two-stage decay process that has its natural explanation by existence of the magnon supercurrent.

To further understand the observed density evolution we consider  three groups of magnons: The magnon BEC group $N_\mathrm{c}(t)$, the nearby ``bottom magnons'' group $N_\mathrm{b}(t)$ and the ``gaseous magnons'' group $N_\mathrm{g}(t)$ occupying the remaining part of the $(\omega,q)$-plane (see Fig.\,\ref{F:1}).
The density dynamics of these magnons may be studied using the phenomenological model developed in Ref.\,\cite{Bozhko2016}.
The model uses the same magnon relaxation frequencies $\gamma_\mathrm{g} = \gamma_\mathrm{b} = \gamma_\mathrm{c}$ for all three groups, a phenomenological parameter $N_\mathrm{cr}$ representing the threshold of Bose-Einstein condensation, and a supercurrent term describing the outflow of coherent magnons from the hot spot with temperature $T$ to the cold part of the film with temperature $T_0$. It is assumed that the supercurrent is driven by a phase difference in the BEC wavefunction, created by the thermally induced change of the saturation magnetization that leads to a frequency shift $\delta\omega_\mathrm{c}(T) = \omega_\mathrm{min}(T)-\omega_\mathrm{min}(T_0)$ between the hot and cold parts of the magnon condensate. In the current work, the presence of the KI process is taken into account by a tenfold increase of the initial density in the ``bottom magnons'' group $N_\mathrm{b}(t)$ in accordance with the experimental BLS data shown in Fig.\,\ref{F:3}(a).
The simulation results presented in Fig.\,\ref{F:3}(c,d) clearly reproduce two distinct stages of the $N_{\mathrm{tot}}(t)$ evolution in the hot spot: A fast initial decay followed by a slower gradual decrease. The density dynamics is in semi-quantitative agreement with the corresponding experimental data. Therefore, we firmly associate the enhancement of the initial decay, observed in our KI experiments, with the temperature induced supercurrent of the magnon BEC \cite{Bozhko2016}. \looseness=-1

To summarize, we showed that parametric pumping of magnons in YIG films is able to create a magnon BEC in a wide range of bias magnetic fields. At the same time, we found that the formation of the magnon condensate is significantly intensified when kinetic instability processes are allowed. 
A narrow intense peak in the population of stochastic magnons, created via this process near the bottom of the magnon spectrum, serves as an efficient precursor for the BEC formation. 
Consequently, the resulting BEC state is denser by an order of magnitude compared to the one created under conditions when only Kolmogorov-Zakharov cascade spectral transfer is allowed. The existence of the condensed magnon state was evidenced by our observation of a two-stage decay of a BEC-related BLS signal after the termination of the parametric pumping. Furthermore, a magnon supercurrent, responsible for this two-stage decay, is stronger in the KI case. We assert that the KI process, being a general physical phenomenon inherent for the systems of non-linear waves, may be found in overpopulated gases of bosonic quasiparticles of different nature, opening thus novel directions of research. \looseness=-1

Financial support by the European Research Council within the Advanced Grant 694709 ``SuperMagnonics'' and by Deutsche For\-schungs\-gemein\-schaft (DFG) within the Transregional Collaborative Research Center SFB/TR 49 ``Condensed Matter Systems with Variable Many-Body Interactions'' as well as by the DFG project INST 248/178-1 is gratefully acknowledged.


\end{document}